\documentclass[12pt]{article}
\begin{document}
\newcommand{\ben}{\begin{enumerate}}
\newcommand{\een}{\end{enumerate}}
\newcommand{\be}{\begin{equation}}
\newcommand{\ee}{\end{equation}}
\newcommand{\bse}{\begin{subequation}}
\newcommand{\ese}{\end{subequation}}
\newcommand{\bea}{\begin{eqnarray}}
\newcommand{\eea}{\end{eqnarray}}
\newcommand{\bc}{\begin{center}}
\newcommand{\ec}{\end{center}}
\newcommand{\mb}{\mbox{\ }}
\newcommand{\vs}{\vspace}
\newcommand{\ra}{\rightarrow}
\newcommand{\la}{\leftarrow}
\newcommand{\IR}{\mbox{I \hspace{-0.2cm}R}}
\newcommand{\IN}{\mbox{I \hspace{-0.2cm}N}}
\newcommand{\ol}{\overline}
\newcommand{\ul}{\underline}
\newtheorem{proposition}{Proposition}
\def\algebra{\Omega_{\xi}({\cal A})}
\def\malgebra{\Omega_{\xi}^{m} ({\cal A})}
\def\evenalgebra{\Omega_{\xi}^{2k}({\cal A})}
\def\oddalgebra{\Omega_{\xi}^{2k+1}({\cal A})}
\author {V. Abramov, N. Bazunova \\  
\\  
        {\it  Institute of Pure Mathematics, University of Tartu,}\\
        {\it  Vanemuise 46, Tartu 51 014, ESTONIA} }

\title{Algebra of differential forms with exterior differential
${d^3=0}$ in dimension one}

\maketitle

\begin{abstract}
In this work, we construct the algebra of differential forms
with exterior differential $d$ satisfying $d^3=0$ on
one-dimensional space. We prove that this algebra is a
graded $q$-differential algebra where $q$ is a cubic root of
unity. Since $d^2\ne 0$ the algebra of differential forms is
generated not only by the first order differential $d x$ but
also by the second order differential $d^2 x$ of a
coordinate $x$. We study the bimodule generated by this
second order differential, and show that its structure is
similar to the structure of bimodule generated by the first
order differential $d x$ in the case of the anyonic line.
\end{abstract}

\section{Introduction}
It is well known that one of possible ways to generalize a
classical Grassmann algebra is to increase the power of
nilpotency of its generators. This means that a possible
generalization can be defined as an associative unital
algebra generated by $\theta_1,\theta_2,\ldots,\theta_n$
satisfying $\theta_i^N=0, N>2, i=1,2,\ldots,n$. In order to
obtain an analogue of classical Grassmann algebra one should
add commutation relations between generators to the above
mentioned definition and this leads to different
generalizations of classical Grassmann algebra in the case
$n>1$. If one imposes the commutation relations
$\theta_i\theta_j=q\;\theta_j\theta_i\, (i<j)$, where $q$ is
an $N$-th primitive root of unity, then the corresponding
structure is called a {\it generalized Grassmann algebra}
\cite{generalized Grassmann algebra},\cite{ternary Grassmann
algebra 2}. Another approach based on ternary commutation
relations and the representation of group of cyclic
permutations $Z_3$ by cubic roots of unity was developed in
\cite{ternary Grassmann algebra 1},\cite{ternary Grassmann
algebra 2} and the corresponding structure is called a {\it
ternary Grassmann algebra}. It should be mentioned that both
generalizations of classical Grassmann algebra mentioned
above coincide in the trivial case of one generator $\theta$
satisfying $\theta^N=0$ and this algebra known as anyonic
line \cite{anyonic line} is closely related to fractional
supersymmetry \cite{fractional supersymmetry}.

A classical Grassmann algebra underlies an exterior
calculus on a smooth manifold with exterior differential
$d$ satisfying $d^2=0$. Therefore the above mentioned
generalizations of Grassmann algebra raise a natural
question of possible generalizations of classical exterior
calculus to one with exterior differential satisfying
$d^N=0,\; N>2$. From an algebraic point of view an
adequate algebraic structure underlying an exterior
calculus is the notion of graded differential algebra.
Hence one can generalize a classical exterior calculus
with the help of an appropriate generalization of graded
differential algebra. This generalization called {\it
graded $q$-differential algebra} was proposed and studied
by M. Dubois-Violette in the series of papers \cite{graded
q-differential algebra}, where the author constructed
several realizations of graded $q$-differential algebra.

According to the definition given by M. Dubois-Violette a
graded $q$-differential algebra is an associative unital
${\bf{N}}$-graded algebra endowed with linear endomorphism
$d$ of degree 1 satisfying $d^N=0$ and the graded
$q$-Leibniz rule
\be
d(\alpha\beta)=d(\alpha)\,\beta+q^{a}\;\alpha\,d(\beta),
\label{q-Leibniz 1}
\ee
where $a$ is the grading of an element $\alpha$, $q$ is a
primitive $N$-th root of unity.

From a point of view of differential geometry the above
definition can be used to generalize the de Rham complex on
a finite-dimensional smooth manifold. This question was
studied in the series of papers \cite{generalized forms 1},
\cite{generalized forms 2}, where the authors used the
notion of ternary Grassmann algebra (considering the first
non-trivial generalization of classical exterior calculus
corresponding to $N=3$) to construct the algebra of
differential forms. If one assumes $d^2\ne 0$ replacing it
by $d^3=0$, then in order to construct a self-consistent
theory of differential forms it is necessary to add to the
first order differentials of local coordinates $dx^1, dx^2,
\ldots, dx^m$ a set of {\it second order differentials} $d^2
x^1, d^2 x^2, \ldots, d^2 x^m$ (and higher order
differentials in the case of $N>3$). Appearance of higher
order differentials, which are missing in classical exterior
calculus, is a peculiar property of a proposed
generalization of differential forms. This has as a
consequence certain problems. It is well known that in
classical exterior calculus functions commute with
differentials, i.e.
\be
f\,dx^i=dx^i\,f,\qquad \forall i=1,2,\ldots,m,
\label{classical}
\ee
where $f$ is a smooth function on a manifold, or, from an
algebraic point of view, the space of 1-forms is a free
finite bimodule over the algebra of smooth functions
generated by the first order differentials and
(\ref{classical}) shows how its left and right structures
are related to each other. Now, assuming that $d$ is no more
classical exterior differential, i.e. $d^2\ne 0$, and
differentiating (\ref{classical}) with regard to graded
$q$-Leibniz rule (\ref{q-Leibniz 1}) one immediately obtains
\be
d^2 x^i\,f=f\,d^2 x^i+[d f,dx^i]_q,
\label{second order differentials}
\ee
where $[d f,dx^i]_q=df\,dx^i-q\;dx^i\,d f$ and we assign
grade 1 to first order differentials $dx^i$. The above
relations (\ref{second order differentials}) are {\it not
homogeneous} in the sense that the commutation relations
between functions and second order differentials include
first order differentials as well.

In this paper, we show that commutation relations between
functions and second order differentials can be made
homogeneous, i.e. they will not include first order
differentials, if we take a non-commutative geometry point
of view. In order to make our construction more transparent
we begin with the simplest case of one-dimensional space. We
construct the differential forms on this space with exterior
differential satisfying $d^3=0$ and show that they form a
graded $q$-differential algebra. In our construction we use
a {\it coordinate calculi} developed in \cite{coordinate
calculi}. Then we study the commutation relations between
functions and second order differentials and show that the
requirement of homogeneity (i.e. vanishing of the second
term in the right-hand side of (\ref{second order
differentials})) implies that a coordinate calculi we are
considering is the differential calculus on the anyonic
line, which means that the commutation relations between
functions and second order differentials are homogeneous,
i.e they take on the form $d^2 x\,f=f\,d^2 x$, only in the
case of the anyonic line (in dimension one).

\section{Graded $q$-differential algebra on one-\\ dimensional space}
In this section, we construct a graded $q$-differential
algebra of differential forms with exterior differential $d$
satisfying $d^3=0$ in dimension one. We study the structure
of a bimodule of second order differentials and show that it
is homogeneous in the case of the anyonic line. In this
section, $q$ is a primitive cube root of unity, i.e.
$q^3=1$.

Let $\cal{A}$ be a free unital associative $\bf C$-algebra
generated by a variable $x$. If $\xi:{\cal A}\to {\cal A}$
is a homomorphism of this algebra and $\partial:{\cal A}\to
{\cal A}$ is a linear map such that
\be
\partial(x)=1,\qquad
\partial(f\,g)=\partial(f)\,g+\xi(f)\,\partial(g),\quad
   \forall f,g\in {\cal A},
\label{partial Leibniz}
\ee
then according to coordinate calculi the map
\be
d:f\to \partial(f)\,d x,
\label{differential}
\ee
where $d x$ is the first order differential of a variable
$x$, is a coordinate differential, i.e. $d$ is a linear map
$d:{\cal A}\to D_{\xi} (A)$ satisfying the Leibniz rule
\be
d(f\,g)=d(f)\,g+f\,d(g),
\ee
and $D_{\xi} (A)$ is a free left module over $\cal A$
generated by $d x$ with the right module structure defined
by the commutation rule
\be
dx\,f=\xi(f)\,d x.
\label{function}
\ee
If $f=\sum_m \alpha_m\,x^m$ is an element of $\cal A$ then
the derivative $\partial:{\cal A}\to {\cal A}$ can be
written in explicit form
\be
\partial(f)=\sum_{m\geq 1}\,\sum_{k=0}^{m-1}\;\alpha_m\,
\xi^k(x)\,x^{m-k-1}.
\label{derivative}
\ee
In order to construct a generalization of exterior calculus
with exterior differential $d$ satisfying $d^2\neq 0$ let us
introduce the second order differential $d^2 x$. Let $(d
x)^k\,(d^2 x)^m$ be monomials composed from first and second
order differentials, where $k,m$ nonnegative integers. As
usual we assume that $(d x)^0=(d^2 x)^0=1$ where $1$ is the
unit element of $\cal A$. Let $\algebra$ be a free left
module over the algebra $\cal A$ generated by the above
introduced monomials. It is easy to see that ${\cal
A}\subset \algebra,\; D_{\xi}({\cal A})\subset \algebra$.
The module $\algebra$ becomes an unital associative algebra
if we define a multiplication law on $\algebra$ by the
relations
\begin{eqnarray}
dx\,x &=& \xi(x)\,d x,\;\;\;\qquad
   d^2 x\,f=\xi(f)\,d^2 x +
   [\partial,\xi]_q\,(f)\,(d x)^2,\label{commutation relations I}\\
(d x)^3 &=& 0,\qquad\;\;\;\;\;\;\;\;\;\;\;\;
    d^2 x\,d x = q^2\,dx\,d^2 x,
    \label{commutation relations II}
\end{eqnarray}
where
$[\partial,\xi]_q(f)=\partial\,(\xi(f))-q\,\xi\,(\partial(f))$,
and $f\in {\cal A}$.

Analyzing the defining commutation relations
(\ref{commutation relations I},\ref{commutation relations
II}) of the algebra $\algebra$ one can note that the
bimodule $D_{\xi}({\cal A})$ of a coordinate calculi is a
submodule of $\algebra$ and the relation (\ref{function})
between its left and right structures follows from the
commutation relation between the first order differential
$dx$ and a variable $x$. The second remark concerns the
structure of the algebra $\algebra$ with respect to the
second order differential. The relations (\ref{commutation
relations I}, \ref{commutation relations II}) show that any
power of the second order differential does not vanish.
Hence the algebra $\algebra $ is an infinite-dimensional
vector space and an arbitrary element $\omega$ of this
algebra can be written in the form
\be
\omega=\sum_{m\geq 0}\sum_{k=0}^{2}\,f_{km}\,(dx)^k\,(d^2
x)^m,\qquad f_{km}\in {\cal A}.
\label{arbitrary element}
\ee
We shall call elements of the algebra $\algebra$
differential forms on one-dimensional space generated by a
variable $x$. The algebra of differential forms $\algebra$
becomes a $\bf N$-graded algebra if we assign grading zero
to each element of the algebra $\cal A$ and grading $k+2m$
to monomial $(dx)^k(d^2 x)^m$, i.e. we assume that a
variable $x$ has grading zero and the gradings of the
differentials $dx,d^2 x$ are respectively 1,2. Then the
algebra of differential forms splits into the direct sum of
its subspaces
$$
\algebra=\bigoplus_{m=0}^{\infty}\;\malgebra
$$
where
\begin{eqnarray}
\Omega_{\xi}^0({\cal A}) &=& {\cal A},\nonumber\\
\evenalgebra &=& \{f\,(d^2 x)^k+h\,(dx)^2\,(d^2
      x)^{k-1}:\;f,h\in {\cal A}\},\; k=1,2,\ldots
      \label{evenalgebra}\\
\oddalgebra &=& \{ f\,dx\,(d^2 x)^k: f\in {\cal A}\},\quad
k=0,1,\ldots.\label{oddalgebra}
\end{eqnarray}
\\
\\
We now extend the differential (\ref{differential}) of the
coordinate calculi to the whole algebra $\algebra$ as
follows:
\vskip.3cm
\begin{tabular}{l l l l l l l}
& & & & & $d(\omega)=(\partial\,f - h)\,dx\,(d^2 x)^{k}$,&
     $\omega\in \evenalgebra$\\
& & & & & & \\ & & & & & $d(\omega)=f\,(d^2 x)^{k+1}
+\partial f\,(dx)^2\,(d^2 x)^{k}$,
        &$\omega\in \oddalgebra$.
\end{tabular}
\\
\\
\\
We shall call the above defined differential an exterior
differential on the algebra of differential forms
$\algebra$. It follows from the definition that exterior
differential is an endomorphism of degree 1 of the algebra
$\algebra$, i.e. $d:\malgebra\to \Omega_{\xi}^{m+1}$.
\begin{proposition}\label{prop:graded algebra}
The algebra of differential forms $\algebra$ is a graded
$q$-differential algebra with respect to exterior
differential $d$, i.e. for any two differential forms
$\omega, \theta$ the exterior differential $d$ satisfies
\begin{eqnarray}
d^3(\omega)&=&0,\label{cube nilpotency}\\
d(\omega\theta)&=&
      d(\omega)\,\theta+q^{\vert
      \omega\vert}\;\omega\,d(\theta),
      \label{q-Leibniz 2}
\end{eqnarray}
where ${\vert\omega\vert}$ is the grading of a form
$\omega$.
\end{proposition}
{\bf Proof.} It follows from
(\ref{evenalgebra},\ref{oddalgebra}) that any differential
form $\omega$ can be decomposed into the sum of two forms
$\omega_o, \omega_e$ respectively of odd and even grading,
where
\bea
\omega_e &=& \sum_{k\geq 1}\,[f_{k}\,(d^2 x)^k + h_{k-1}\,(d x)^2\,(d^2
x)^{k-1}],\label{evenform}\\
\omega_o &=& \sum_{k\geq 0}\,g_k\,dx\,(d^2 x)^k.\nonumber
\eea
From the definition of the exterior differential it also
follows that a differential form of even
grading(\ref{evenform}) is $d$-closed ($d\omega_e=0$) if it
satisfies the following condition
\be
\partial f_{k}=h_{k-1}.
\label{closed}
\ee
Now it is easy to show that any form of odd grading is
$d^2$-closed, i.e. $d^2 \omega_o=0$. Indeed applying the
exterior differential $d$ to form $\omega_o$, one obtains
the form of even grading
$$
d\omega_o=\sum_{k\geq 0} [\,g_k\,(d^2 x)^{k+1}+\partial
g_k\,(d x)^2\,(d^2 x)^k],
$$
which is $d$-closed according to (\ref{closed}).
Differentiating form (\ref{evenform}) of even grading twice,
one obtains the form
\be
d^2 \omega_e=\sum_{k\geq 1}[\, (\partial f_k-h_{k-1})\,(d^2
x)^{k+2}
    +(\partial^2f_k-\partial h_{k-1})\,(d x)^2\,(d^2 x)^{k+1}],
\ee
which is $d$-closed. Thus the cube nilpotency (\ref{cube
nilpotency}) of the exterior differential is proved. The
$q$-Leibniz rule (\ref{q-Leibniz 2}) can be verified by a
direct calculation.

A homomorphism $\xi$ plays a role of a parameter in the
structure of the algebra of differential forms, and,
choosing particular homomorphism, we can specify the
structure of $\algebra$. We remind that according to the
definition the algebra of differential forms $\algebra$ is a
free left module over the algebra ${\cal A}$ generated by
the monomials $(d x)^k (d^2 x)^m$ with associative
multiplication law determined by the relations
(\ref{commutation relations I}, \ref{commutation relations
II}). Actually this algebra is generated by three generators
$x, d x, d^2 x$. Hence its structure will be more
transparent if we define it by means of {\it commutation
relations} imposed on the generators $x, d x, d^2 x$. The
only relation in (\ref{commutation relations I},
\ref{commutation relations II}), which is not a commutation
relation between generators, is the second relation in
(\ref{commutation relations I}) containing an arbitrary
element $f$ of the algebra $\cal A$. The reason why it
contains an arbitrary element $f$ is that in contrast to the
commutation relation $dx\, x=\xi(x)\,d x$ the right-hand
side of this relation is not a homomorphism of the algebra
$\cal A$ because of the non-homogeneous term
$[\partial,\xi]_q(f)$. Obviously imposing the condition
\be
[\partial,\xi]_q(f)=0,
\label{condition}
\ee
we can replace the second relation in (\ref{commutation
relations I}) by the commutation relation
\be
d^2 x\, x=\xi(x)\,d^2 x,
\ee
which has exactly the same form as the first one involving
the first order differential.

Actually the choice for a homomorphism $\xi$ is not very
wide. Indeed every homomorphism $\xi$ of the algebra $\cal
A$ is determined by an element $h_{\xi}\in {\cal A}$ such
that $\xi(x)=h_{\xi}$. From (\ref{partial Leibniz}) it
follows that if the derivative $\partial$ determined by a
homomorphism $\xi$ should satisfy $\partial(x^m)\sim
x^{m-1}$ then $\xi(x)=\alpha_{\xi}\,x$ where $\alpha_{\xi}$
is a complex number. The condition (\ref{condition}) can be
solved with regard to a homomorphism $\xi$, and the
following proposition describes a structure which is induced
on one-dimensional space in this case.

\begin{proposition}\label{homomorphism}
The condition (\ref{condition}) is satisfied if and only if
$\xi(x)=q\,x$. This solution leads to the $q$-differential
calculus on anyonic line with derivative
\be
\partial(f)=\sum_{k\geq 1}\,
\alpha_k\,\frac{x^{k-1}}{[k-1]_q!},\quad f=\sum_{k\geq
0}\,\alpha_k\,\frac{x^{k}}{[k]_q!}
\label{q-derivative}
\ee
This means that one can consistently add the relation
$x^3=0$ to the relations (\ref{commutation relations I},
\ref{commutation relations II}).
\end{proposition}
{\bf Proof}. Using the formula (\ref{derivative}) one can
find
\bea
\partial(\xi(f)) &=& \alpha_{\xi}\,\sum_{m}\,\alpha_{m}
   \sum_{k=0}^{m-1}\,\xi^k(h)\,h^{m-k-1},\nonumber\\
q\, \xi(\partial(f)) &=& q\,\sum_{m}\,\alpha_{m}
   \sum_{k=0}^{m-1}\,\xi^k(h)\,h^{m-k-1}.\nonumber
\eea
Thus
\be
[\partial,\xi]_q(f)=(\alpha_{\xi} - q)\,\sum_{m}\,\alpha_{m}
   \sum_{k=0}^{m-1}\,\xi^k(h)\,h^{m-k-1}.
\ee
From the above formula it immediately follows that
$\xi(x)=q\,x$ or $\xi(f(x))=f(q x)$. Putting $\xi(x)=q\,x$
in (\ref{derivative}), one obtains (\ref{q-derivative}). It
was explained in \cite{fractional supersymmetry} that
$q$-differential calculus determined by the derivative
(\ref{q-derivative}) is correctly defined at cube root of
unity on one-dimensional space generated by a variable $x$
only in the case when $x^3=0$. It is easy to show that the
relation $x^3=0$ can be consistently added to the relations
(\ref{commutation relations I}, \ref{commutation relations
II}).

Now the algebra of differential forms $\algebra$ on anyonic
line can be defined as an unital associative algebra
generated by three generators $x, d x, d^2 x$ satisfying the
following commutation relations:
\begin{eqnarray}
x^3 &=& 0,\cr dx\,x &=& q\,x\,d
x,\qquad\qquad\qquad\;\;\;\;\;
   d^2 x\,x=q\, x\,d^2 x \label{q-commutation relations I}\\
(d x)^3 &=& 0,\qquad\qquad\qquad\;\;\;\;\;\;\;\;\;\;
    d^2 x\,d x = q^2\,dx\,d^2 x.
    \label{q-commutation relations II}
\end{eqnarray}

\section*{Acknowledgments}
The authors would like to acknowledge the financial support
of Estonian Science Foundation under the grants No.3308 and
No.1134.

\end{document}